\documentclass[]{aa}

\newcommand{\xsource}{4U 1820-30\xspace}
\usepackage{natbib,twoopt}
\usepackage{multirow}
\usepackage[breaklinks=true]{hyperref}
\usepackage{amsmath}% Advanced maths commands
\usepackage{amssymb}	% Extra maths symbols
\usepackage[T1]{fontenc}
\usepackage{booktabs}
\usepackage{threeparttable}
\usepackage{xspace}
\usepackage{subcaption}
\usepackage{array} % Assicurati di includere questo pacchetto nel preambolo
\usepackage{graphicx} % Necessario per il ridimensionamento
\usepackage{booktabs} % Migliora lo stile delle tabelle, opzionale
\usepackage{tabularx}
\usepackage{graphicx}
\usepackage{txfonts}
\bibpunct{(}{)}{;}{a}{}{,} % to follow the A&A style
\usepackage{siunitx}
\newcommand{\rg}{\ensuremath{\rm{R_{g}}} \xspace}

\newcommand{\ixpe}{\textit{ixpe}}
\newcommand{\chisqr}{\ensuremath{\chi^{2}}\xspace}

\begin{document} 

\title{Unveiling the reflection spectrum in the ultracompact LMXB 4U 1820-30}

 %\subtitle{}

   \author{
   A. Anitra
   \inst{\ref{in:Uni_Cagliari},\ref{in:Uni_Palermo},\thanks{\email{alessio.anitra@inaf.it}}},
   A. Gnarini
   \inst{\ref{in:UniRoma3}}, 
   T. Di Salvo
   \inst{\ref{in:Uni_Palermo}}, 
   R. Iaria
   \inst{\ref{in:Uni_Palermo}}, 
   A. Sanna
   \inst{\ref{in:Uni_Cagliari}}, 
   L. Burderi
   \inst{\ref{in:Uni_Cagliari},\ref{in:IASF},\ref{in:Inaf_Cagliari}},
   A. Marino
   \inst{\ref{in:IASF},\ref{ICE_CSIC},\ref{IEEC}}, 
   M. Del Santo
   \inst{\ref{in:IASF}}, 
   G. Matt
   \inst{\ref{in:UniRoma3}},
   F. Ursini
   \inst{\ref{in:UniRoma3}}, 
   S. Bianchi
   \inst{\ref{in:UniRoma3}}, 
   F. Capitanio
   \inst{\ref{in:INAF-IAPS}}, 
   S. Fabiani
   \inst{\ref{in:INAF-IAPS}}, 
   A. Tarana
   \inst{\ref{in:INAF-IAPS}},
   A. Di Marco
   \inst{\ref{in:INAF-IAPS}}   
   }

\institute{
Dipartimento di Fisica, Universit\`a degli Studi di Cagliari, SP
Monserrato-Sestu, KM 0.7, Monserrato, 09042 Italy \label{in:Uni_Cagliari}
\and
Dipartimento di Fisica e Chimica - Emilio Segrè,
Universit\`a di Palermo, via Archirafi 36 - 90123 Palermo, Italy \label{in:Uni_Palermo}
\and
Dipartimento di Matematica e Fisica, Università degli Studi Roma Tre, via della Vasca Navale 84, I-00146 Roma, Italy \label{in:UniRoma3}
\and
INAF/IASF Palermo, via Ugo La Malfa 153, I-90146 Palermo, Italy \label{in:IASF}
\and
INFN, Sezione di Cagliari, Cittadella Universitaria, 09042 Monserrato, CA, Italy \label{in:Inaf_Cagliari}
\and
Institute of Space Sciences (ICE, CSIC),Campus UAB, Carrer de Can Magrans s/n, E-08193 Barcelona, Spain \label{ICE_CSIC}
\and
Institut d'Estudis Espacials de Catalunya (IEEC), Esteve Terradas 1, RDIT Building, Of. 212 Mediterranean Technology Park (PMT), 08860, Castelldefels, Spain \label{IEEC}
\and
Department of Physics, University of Trento, Via Sommarive 14, 38122 Povo (TN), Italy \label{in:Treno_uni}
\and
INAF -- Istituto di Astrofisica e Planetologia Spaziali, Via del Fosso del Cavaliere 100, 00133 Roma, Italy \label{in:INAF-IAPS}
}

  \abstract
   {4U 1820–30 is a well-known ultracompact X-ray binary located in the globular cluster NGC 6624, consisting of a neutron star accreting material from a helium white dwarf companion characterized by the shortest known orbital period for this type of star (11.4 minutes). Despite extensive studies, the detection of the relativistic Fe K emission line, indicative of a Compton reflection component, has been inconsistently reported and no measurement of the system inclination has been achieved (although it is hypothesized to be low).
In this work, we investigate the broadband spectral and polarimetric properties of 4U 1820–30, exploring the presence of a reflection component and its role in shaping the polarization signal. 

We analyzed simultaneous X-ray observations from NICER, \textit{NuSTAR}, and IXPE. 
The spectral continuum was modeled with a disk blackbody, a power law, and a Comptonization component with seed photons originating from a boundary layer.
We detected a strong reflection component,  described, for the first time, with two self-consistent models ({\tt Relxillns} and {\tt Rfxconv}), allowing us to provide a measurement of the system inclination angle ($31^{+9}_{-5} \, {\rm degrees}$), supporting the low-inclination hypothesis. 
Subsolar iron abundances were detected in the accretion disk and interstellar medium, probably related to the source location in a metal-poor globular cluster.

The source exhibits an increase in polarization  from an upper limit of $1.2\%$ in the 2--4 keV band up to about $8\%$ in the 7--8 keV range. The
main contribution to the polarization comes from the Comptonization, with a possible significant contribution from the weak, but highly polarized reflection component. The disk is expected to be orthogonally polarized to these components, which may help to explain the decreasing of the observed polarization at low energies. However, the high polarization degree we found in the 7--8 keV band challenges the current models, also taking into consideration  the relatively low inclination angle derived from the spectral analysis. 

}
 
  \authorrunning{A. Anitra et al.}

  \titlerunning{Unveiling the reflection spectrum in the ultracompact LMXB 4U 1820-30}
  
  
% \object{X 1822-371}

   \maketitle
%
%________________________________________________________________
\section{Introduction}

The accretion flow around neutron stars (NSs) in low-mass X-ray binaries \citep[see][for a review]{DiSalvo_2023hxga.book..147D} typically includes an optically thick accretion disk, a boundary layer connecting the disk to the NS's surface, and a cloud of hot electrons called the "corona." These components contribute  to the X-ray spectrum in different ways, with the disk emitting as a multi-temperature blackbody, while the electrons in corona upscatter the soft photons, resulting in a high-energy (cutoff) power-law shape component. The boundary layer can also directly contribute to a blackbody component or act as a seed-photon source for the Comptonization component \citep{Mondal_2020MNRAS.494.3177M,Marino_2023MNRAS.525.2366M}.

Among NS low-mass X-ray binaries (LMXBs), 4U 1820-30 stands out due to its unique characteristics. The system hosts a NS accreting material from a white dwarf companion and features an ultrashort orbital period of just 11.4 minutes, classifying it as an ultracompact X-ray binary \citep[UCXB;][]{Stella_1987ApJ...315L..49S}.
It is located in the NGC 6624 globular cluster, approximately 8 kpc away \citep{Baumgardt_2021MNRAS.505.5957B}. 
4U 1820-30 is categorized as an "atoll" source,
observed in both soft and hard spectral states, often termed the "banana" and "island" states, respectively \citep{Hasinger_1987IAUC.4489....3H,Tarana_2007ApJ...654..494T}.

The source is also known for its luminosity modulation, often referred to as super-orbital modulation, characterized by a periodic change (around 170-176 days) in the X-ray luminosity over timescales much longer than the orbital period of the binary. 
It was thought to be stable and attributed to the gravitational influence of a third star, based on the triple model hypothesis \citep{Grindlay_1988AdSpR...8b.539G,Chou_2001ApJ...563..934C}. However, recent studies indicate that the period has decreased from 171 to 167 days over the past 36 years, challenging the triple model and suggesting an irradiation-induced mass transfer instability as the cause \citep{Chou_2024arXiv240908451C}.
As a result, the source alternates between a low-luminosity mode, with $L_{\text{low}} \sim 8 \times 10^{36} \, \text{erg s}^{-1}$, and a high-luminosity mode, with $L_{\text{high}} \sim 6 \times 10^{37} \, \text{erg s}^{-1}$. Variations in luminosity and mass transfer rate can significantly affect the accretion disk structure and the boundary layer, leading to changes in the spectral state and the presence of spectral features, such as the Fe K emission line.
For instance, modulation has been linked to the clustering of Type-I X-ray bursts, which occur more frequently when the source is in the low mode \citep{Titarchuk_2013ApJ...767..160T}.
 
The detection of the relativistic Fe K emission line, indicative of a reflection component, has been inconsistently reported in the literature \citep{Cackett_2008ApJ...674..415C,Titarchuk_2013ApJ...767..160T,Mondal_2016MNRAS.461.1917M}. All the studies often employ non-self-consistent modeling that fails to describe the full reflection spectrum. This feature in LMXBs serves as a powerful diagnostic tool, offering critical insights into the system geometry, the physical properties of the accretion disk, and the behavior of matter in extreme conditions \citep[e.g.,][]{Anitra_2024A&A...690A.148A}.

In this paper, we investigate the system geometry by performing a combined spectral and polarimetric analysis on X-ray data obtained from simultaneous NICER, \textit{NuSTAR}, and IXPE observations (see also \citealt{DiMarco.4U1820}). Polarization is indeed very sensitive to the geometry of the accreting system \citep{Gnarini.etAl.2022,Gnarini.etAl.2024,Bobrikova.etAl.2024} and provides a new approach to understanding the physical properties of X-ray emission from LMXBs. In particular, IXPE has revealed so far that, for most observed sources, the main source of X-ray in NS-LMXBs is the Comptonization in the spreading/boundary layer \citep[see e.g.,][]{Ursini.etAl.2023,Gnarini.etAl.2024}, plus the contribution of reflected photons on the accretion disk. Our study confirms the presence of a strong reflection component in the spectrum and a subsolar iron abundance in both the disk material and the interstellar medium, providing an estimation of the system inclination. The presence of the reflection component is also important for the X-ray polarization because the reflected photons are expected to be highly polarized. 

\section{Data analysis}

4U 1820-30 was observed by the Neutron Star Interior Composition Explorer (NICER) on October 11, 2022 (ObsId 5050300117), and on April 15 (ObsId 6689020101) and April 16, 2023 (ObsId 6689020102), with a total exposure time of 26 ks. We processed NICER data using the {\tt nicerl2} pipeline tool in NICERDAS v12, available with HEASoft v6.34, following the recommended calibration procedures and standard screening (the calibration database version used was {\tt xti20240206}). We accumulated light curves from 0.3 to 10 keV for the three observations using {\tt  nicerl3-lc} to check for Type-I X-ray bursts, finding none. 
To generate the spectrum, we used the {\tt nicerl3-spect} tool, excluding detectors 14 and 34 due to elevated noise levels. The background file "{\tt scorpeon}" was chosen with the {\tt bkgformat=file} option. Following recommendations from the NICER help team, we applied a systematic error as indicated in the latest calibration file and grouped the data using an optimal rebinning ({\tt grouptype=optmin}) and ensuring at least 25 counts per energy bin \citep{Kaastra_2016A&A...587A.151K}.

The Nuclear Spectroscopic Telescope Array (\textit{NuSTAR}), observed the source on October 12, 2022 (ObsID 90802327002) and April 15 and 16, 2023 (ObsIDs 90902308002 and 90902308004), with a total exposure time of 49 ks. The data reduction, as well as spectrum and light curve extraction, was performed using the {\tt nupipeline} and {\tt nuproducts} tools in HEASoft. A circular region with a radius of 150 arcseconds was used to extract source events, while background events were obtained from an equivalent-sized circular region  within the same detector quadrant. We systematically searched for Type-I X-ray bursts in the \textit{NuSTAR} data, but none of them were detected.
As with the NICER spectra, we grouped the \textit{NuSTAR} data using the {\tt ftgrouppha} tool, applying optimal rebinning and ensuring at least 25 counts per energy bin.

The Imaging X-ray Polarimetry Explorer \citep{Weisskopf.etAl.2016,Weisskopf.2022} observed 4U 1820-30 twice on October 11 2022 and between 15 and April 16 2023 (ObsID 02002399), for a total exposure of 102.5 ks per detector unit (DU). To extract the source spectra and light curves, we processed the level 2 data using \textsc{xselect}, considering a circular region of 120 arcseconds centered on the source and the weighted analysis scheme \citep{DiMarco.etAl.2022}. Since the source is bright ($> 2$ counts s$^{-1}$ arcmin$^{-2}$), no background subtraction or rejection was applied \citep{DiMarco.etAl.2023}. We generated the ancillary response file (ARF) and the modulation response file (MRF) with \texttt{ixpecalcarf}. 

\section{Spectral analysis}
We initially performed a simultaneous fit to analyze the spectral properties and evaluate the consistency of the three datasets. We used the energy range of 0.4--10 keV for the NICER spectrum, {as recommended in the calibration guidelines}\footnote{As reported, below 0.4 keV there are several uncertainties that start to compete, like gain scale, RMF parameters and the Carbon edge profile (see \url{https://heasarc.gsfc.nasa.gov/docs/nicer/analysis_threads/cal-recommend/} for more details).}, and 4.5--35 keV for the \textit{NuSTAR} spectrum, excluding energies below 4.5 keV due to a divergence caused by an excess flux in the FPMA camera \citep{Madsen_2020arXiv200500569M}, as well as those over 35 keV, where the background spectrum starts to dominate over the source.

The spectral analysis was conducted with XSPEC v12.14.1 \citep{Arnaud_1996}.
The continuum was modeled using a multi-color blackbody component ({\tt diskbb} in XSPEC \citealt{1986_Makishima}) and a Comptonized component {\tt comptb}. The {\tt comptb} model deviates from other Comptonization models by incorporating a component for inward bulk motion, although this aspect can be neglected by setting the bulk efficiency over the thermal Comptonization, denoted as $\delta$, to zero (see \citealt{Farinelli_2008ApJ...680..602F} for a complete description of the model).
We also fixed the {\tt logA} parameter of the {\tt comptb} component to a value of {\tt 8}. This parameter, known as the illuminating factor, represents the distribution of seed photons, where \(A/(1 + A)\) corresponds to the fraction scattered by the Compton cloud, while \(1/(1 + A)\) is the fraction escaping without scattering. By setting {\tt logA} to the value 8, we assumed that all seed photons were up-scattered.

To account for interstellar medium absorption and to allow for variations in iron and oxygen abundances, we multiplied total emission by the Tuebingen-Boulder ISM absorption model ({\tt TBfeo} in XSPEC). This adjustment was crucial, as NICER spectra often exhibit residuals potentially linked to the oxygen K edge (0.56 keV) and the iron L edge (0.71 keV),  suggesting an under- or overabundance of these elements in the interstellar medium or ionization of neutral oxygen by solar X-ray radiation\footnote{See:\ \url{https://heasarc.gsfc.nasa.gov/docs/nicer/data_analysis/workshops/NICER-CalStatus-Markwardt-2021.pdf}}. We adopted the most recent ISM abundances and cross sections as described in \citet{Wilms_2000ApJ...542..914W} and \citet{Verner_1996ApJ...465..487V}.

When data from different instruments are combined, it is important to account for potential discrepancies in their flux cross-calibrations. To address this, we included a multiplicative constant, fixed at 1 for the first spectrum, and allowed to vary for the others. Thus, our preliminary model was defined as:
${\tt Model\,0: const*TBfeo*(Comptb+diskbb)}$.
\begin{figure}
    \centering
    \includegraphics[width=1\linewidth]{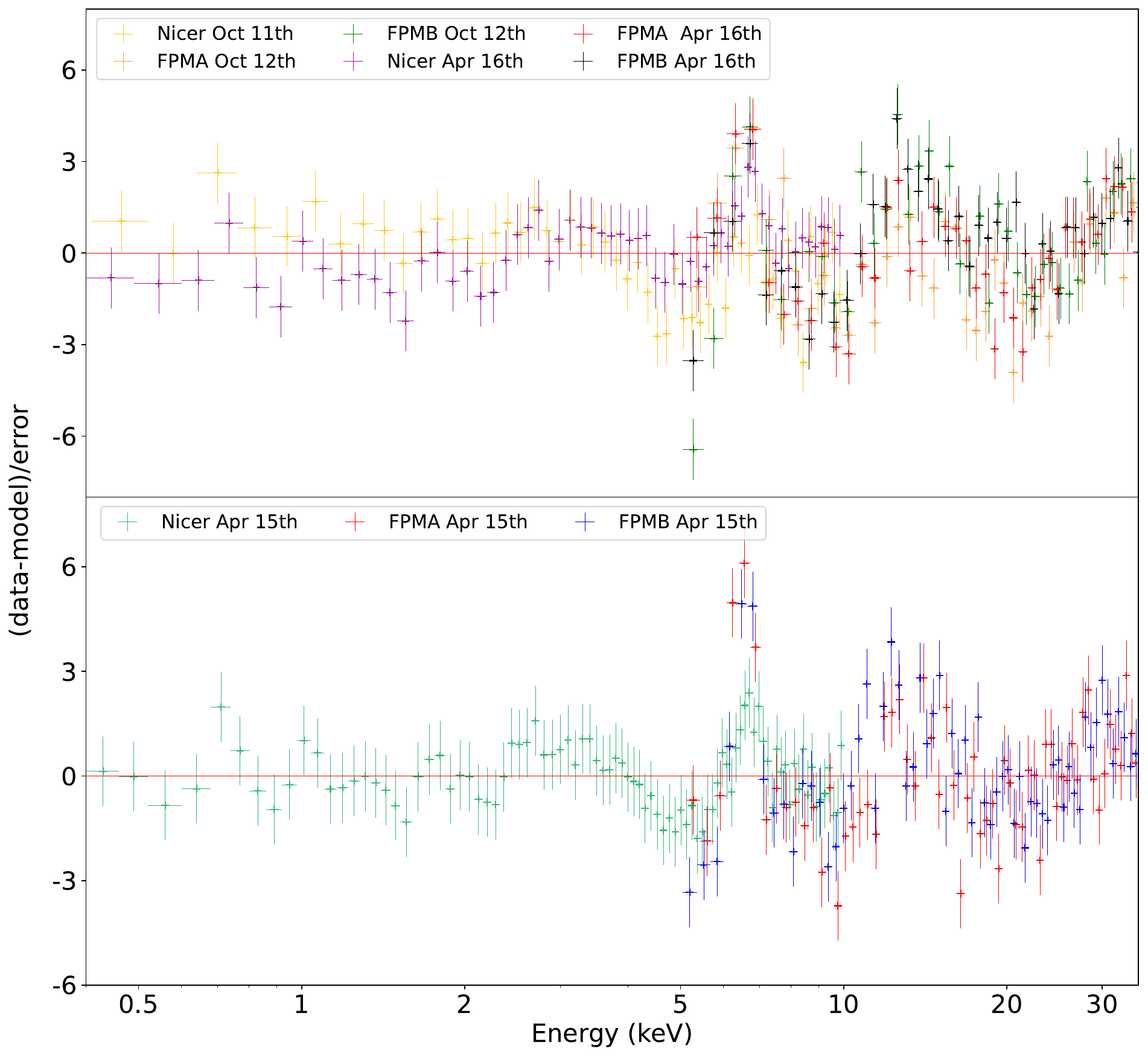}
    \caption{\label{no_gauss} Residuals in units of sigma with respect to {\tt Model 0} for the October and April 16 data (top panel) and April 15 data (bottom panel). Data were rebinned for visual purposes only.}
    \label{fig:enter-label}
\end{figure}

When we performed a simultaneous fitting of the NICER and \textit{NuSTAR} data, a slight offset between the spectral slopes was observed \citep{Ludlam_2020ApJ...895...45L}. To account for this instrumental discrepancy, we left the index of the Comptonization component, $\alpha$ in XSPEC,  untied between NICER and \textit{NuSTAR,}  allowing for some flexibility between the two datasets. Nevertheless, we systematically verified that the difference never exceeded 10$\%$. 

The application of this model showed that the April 15 spectra deviated from the others, suggesting a slightly different spectral state. Consequently, we decided to analyze these data separately while fitting the October and April 16 data simultaneously. {This choice is further justified by the slightly different spectral state of the source across the three observations. As reported in \citealt{DiMarco.4U1820}, while the observations from October 11–12 and April 16 exhibit a comparable hardness ratio (defined as the difference between the counts in the 4–10 keV band and the 0.3--4 keV band, divided by the counts in the 0.3--10 keV band), the April 15 observation shows a higher hardness ratio, indicating a slightly harder spectrum.}

The model fails to fit the entire spectrum, revealing an emission feature around 6.6 keV in both fits. This feature significantly improves the fit when modeled with a Gaussian line at 6.57 keV, likely corresponding to Fe XXV K${\alpha}$ emission line. Additionally, the model exhibits a flux excess at energies above 30 keV (see Fig. \ref{no_gauss}). We initially considered the possibility that this excess might be due to a dominant background contribution in this energy range, but we ruled out this hypothesis, as the background level is not comparable to the observed data in this band.
High-energy excesses such as these have been frequently observed in the literature for various sources \citep[e.g.,][]{Dai_2010A&A...516A..36D}.
This may arise from an additional Comptonization process involving a plasma of electrons with a non-thermal distribution \citep[see][]{DiSalvo_2000ApJ...544L.119D,DiSalvo2006ApJ...649L..91D}. In such a scenario, the seed photon radiation spectrum must originate from the same region as the seed photons for the thermal component. This assumption prevents an overestimation of the hard-tail contribution at lower energies, which could otherwise lead to incorrect modeling of the low-energy continuum.
To simulate this contribution, we indeed introduce a power-law component with a low-energy exponential roll-off at the seed photon Comptonization temperature, modeled using the \texttt{expabs*powlaw} combination. 
 The updated model was, therefore:
{\tt Model 1: const*TBfeo*(expabs*powlaw+Comptb+diskbb+gauss)}.

Introducing the {\tt Gaussian} and {\tt powerlaw} components significantly improved the fit, flattening the residuals and yielding a \chisqr of 788 with 813 degrees of freedom (d.o.f.) for the October and April 16 data and 378 with 418 d.o.f. for the April 15 data.
The best-fit centroid energy of the Gaussian at $6.5 \pm 0.1$ keV and $6.59 \pm 0.06$ keV supports the identification of this feature as arising from the Fe XXV transition. 
Additionally, the high $\sigma$ value ($0.9^{+0.2}_{-0.1}$ keV for the first fit and $0.35^{+0.10}_{-0.08}$ keV for the second) implies that the emission originates from the innermost region of the accretion disk, consistent with an association to an underlying reflection component. % \citep{Titarchuk_2013ApJ...767..160T}.

\subsection{Reflection component}
\begin{table}[t]
\caption{\label{diskline} Best value parameters obtained applying {\tt Model 2} to the data. Uncertainties are calculated at the 90\% confidence level.}
\centering
\renewcommand{\arraystretch}{1.2} % Aumenta l'interlinea (1.5 è un buon valore, ma puoi regolarlo)
\begin{threeparttable}
\begin{tabular}{l c c c}
\hline
Model & Component & Oct + 16 Apr & 15 Apr \\ \hline
\texttt{TBfeo} & nH ($10^{22}$) & $0.168 \pm 0.003$ & $0.170 \pm 0.004$ \\
               & O               & $1.18 \pm 0.05$   & $1.20 \pm 0.07$   \\
               & Fe              & $0.7 \pm 0.3$   & $0.7 \pm 0.4$   \\
               
\texttt{expabs} & ${\rm E_{cut}}$ (keV)  & {\tt ${\rm kT_{s}}$}  \tnote{\dag}        & {\tt ${\rm kT_{s}}$}\tnote{$\dag$}  \\ 
\texttt{powerlaw} & Index     & 2.09\tnote{*}             & 2.34\tnote{*}             \\
                  & Norm          & $0.027 \pm 0.006$ & $0.08 \pm 0.02$ \\ 
                  
\texttt{diskline} & E (keV)  & $6.57 \pm 0.06$   & $6.56^{+0.05}_{-0.07}$   \\
                  & Index       & $-2.5 \pm 0.5$    & $-2.1^{+0.3}_{-0.5}$  \\
                  & ${\rm R_{in}}({\rm R_{g}})$     & <37       &  <52       \\
                  & ${\rm R_{out}}({\rm R_{g}})$ & 1000\tnote{*}          & 1000\tnote{*}               \\
                  & Incl (deg)    & 45\tnote{*}              & 45\tnote{*}              \\
                  & Norm (10$^{-3}$) &   $1.5^{+0.3}_{-0.2}$  & $2.2^{+0.5}_{-0.4}$    \\ 
                  
\texttt{comptb}  & ${\rm kT_{s}}$ (keV)     & $0.84^{+0.05}_{-0.03}$   & $1.00 \pm 0.04$   \\
                  & $\alpha$         & $0.87 \pm 0.02$   & $0.91 \pm 0.03$   \\
                  & ${\rm kT_{e}}$ (keV)     & $2.93 \pm 0.02$   & $2.88 \pm 0.03$   \\
                  & Norm          & $0.055 \pm 0.001$ & $0.070 \pm 0.001$ \\ 
                  
\texttt{diskbb}  & ${\rm T_{in}}$(keV)     & $0.64 \pm 0.03$   & $0.67 \pm 0.03$   \\
                  & Norm          &  $910^{+160}_{-130}$    & $840^{+130}_{-120}$     \\ \hline
                  & $\chi^2/$dof  & 812/813         & 345/416         \\ \hline
\end{tabular}
\begin{tablenotes}
\item[$\dag$] Linked to the photon seed temperature.
\item[*] Kept frozen during the fit.
\end{tablenotes}
\end{threeparttable}
\end{table}
To test for the presence of a reflection component, we replaced the Gaussian component with a non-self-consistent model that describes a line emission from a relativistic accretion disk, using the {\tt diskline} model in XSPEC \citep{Fabian_1989MNRAS.238..729F}.
\begin{figure}
    \centering
    \includegraphics[width=1\linewidth]{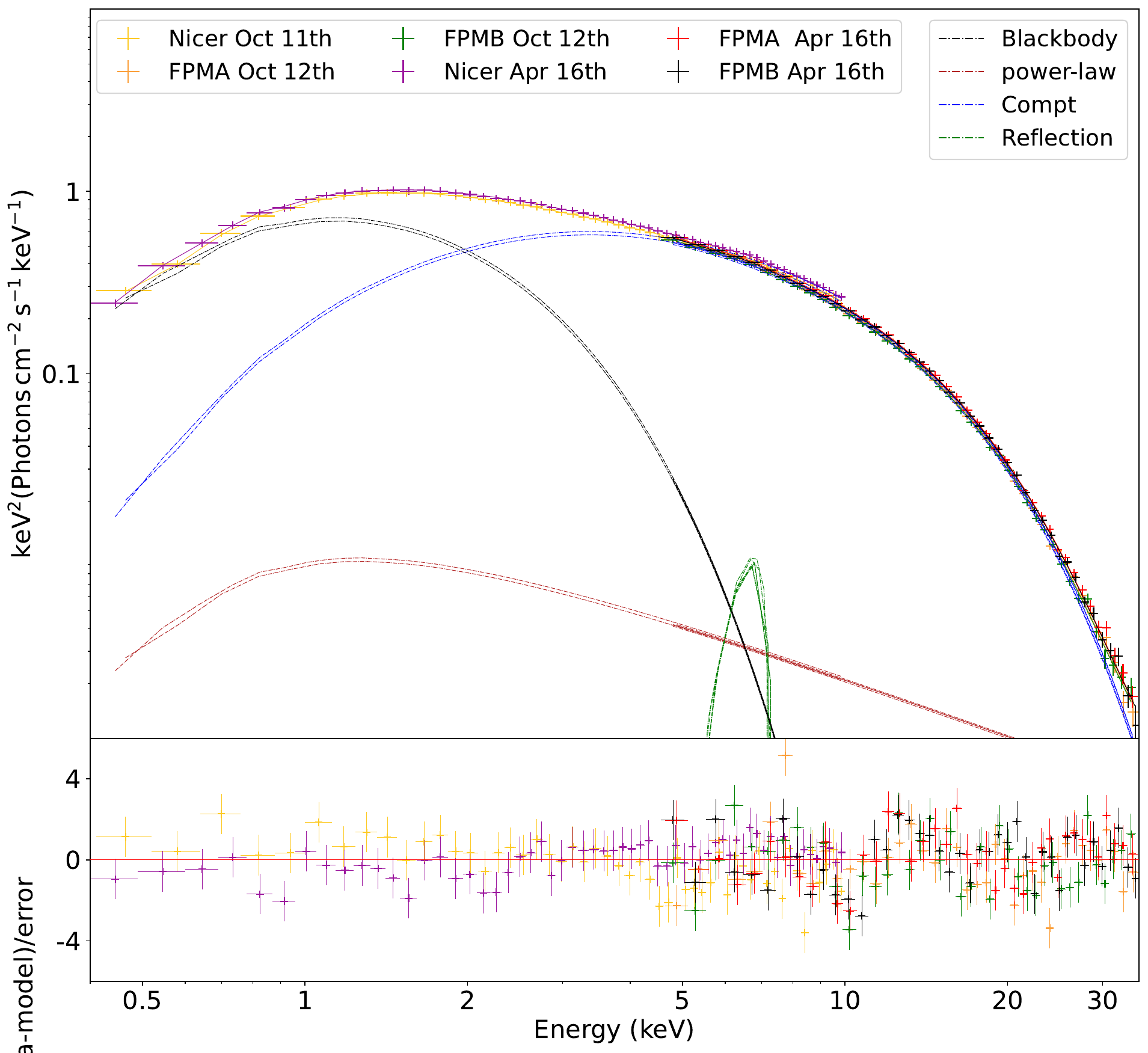} 
    \includegraphics[width=1\linewidth]{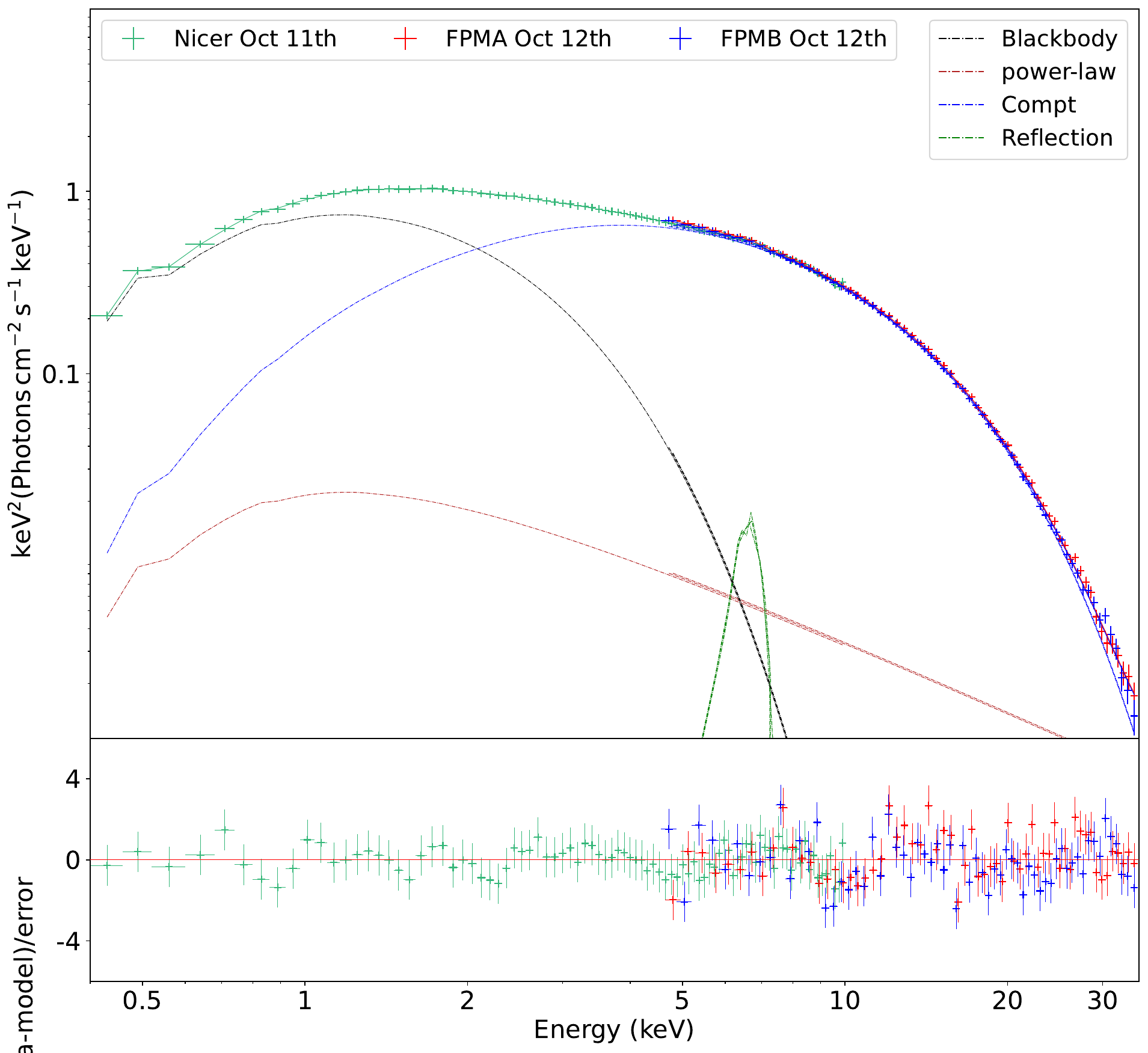} 
    \caption{\label{diskline} Spectra and residuals in units of sigma with respect to {\tt Model 2} for the October and April 16 data (top panel) and April 15  data (bottom panel). Data were rebinned for visual purposes only.}
\end{figure}
The new model ({\tt Model 2}) successfully describes the iron emission line, providing a statistically good \chisqr \ result at 812(813) for the October and April 16 fit and 345(416) for April 15. % compared to {\tt Model 1} in both fits, with a confidence level greater than 7 $\sigma$. 
However, not all the line parameters could be constrained. For instance, the inclination angle of the source was fixed at 45 degrees (consistent with \citealt{Anderson_1997ApJ...482L..69A,Marino_2023MNRAS.525.2366M}).
We believe that this limitation is due to the weakness of the feature, as the normalization of the component is only $(1.6-2)\times10^{-3}$ photon/cm$^2$/s, representing approximately 0.16$\%$ of the total flux. The fit also suggests the line is emitted from a region close to the NS, with an upper limit for the inner radius of 37 \rg  and 52 \rg for the two fits, respectively. The other results are reported in Table \ref{diskline}.

As shown in Fig. \ref{diskline}, the residuals in the $5-10$ keV band were not completely flattened by the addition of the \texttt{diskline} and power-law components. These residuals may arise from additional features, such as the Compton hump beyond 10 keV, pointing to the presence of a reflection continuum in the spectrum. For this reason, we decided to remove the \texttt{diskline} and use a self-consistent reflection model.

We employed the convolution model \texttt{Rfxconv} (developed by \citealt{Kolehmainen_2011MNRAS.416..311K}) to reproduce a reflected spectrum from an ionised disk, applied to a continuum Comptonization model (\texttt{comptb} in XSPEC) with the parameters linked to those of the continuum that have already been included. To account for Doppler and relativistic effects, we used the \texttt{rdblur} convolution model \citep{Fabian_1989MNRAS.238..729F}. 
To describe only the reflection component, we assumed a negative relative reflection fraction ({\tt rel-frac}), allowing it to vary between -1 and 0. The best-fit values for the models are reported in Table \ref{reflection_table}. 

\begin{figure}
    \centering
    \includegraphics[width=1\linewidth]{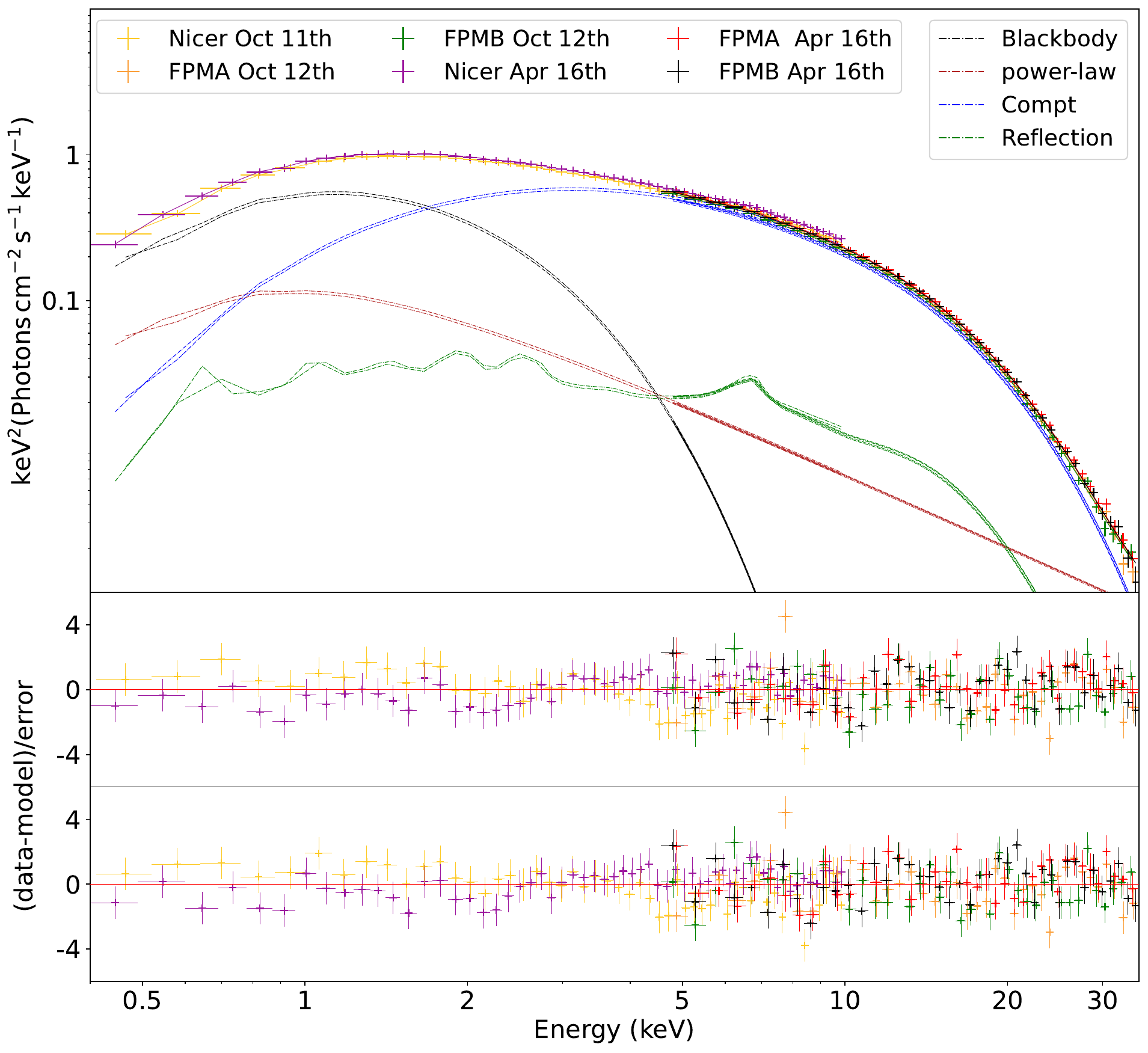} \quad 
    \includegraphics[width=1\linewidth]{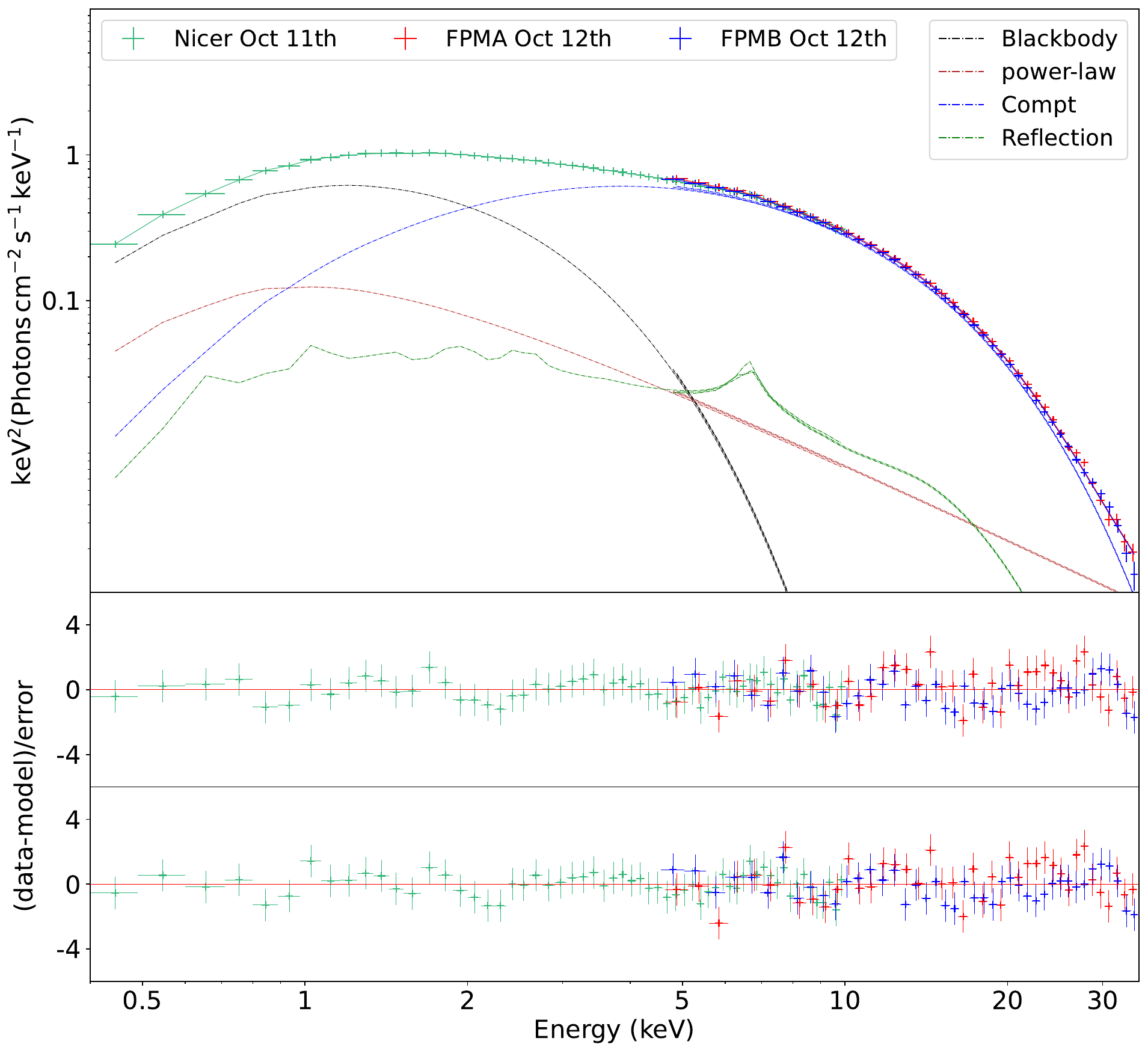} 
    \caption{\label{reflection_plot} Spectra and residuals in units of sigma with respect to {\tt Model 3} (second panel) and {\tt 3A} (third panel) for the October and April 16 data (upper panel) and April 15 data (bottom panel). Data were rebinned for visual purposes only.}
\end{figure}

The flattened residuals, shown in Fig. \ref{reflection_plot}, indicate a significantly improved fit. The application of the new model ({\tt Model 3}) not only reduced the $\chi^2$ values in both fits ($\chi_{oct+apr}^{2}({\rm dof})= 762(810)$ and $\chi_{apr}^{2}({\rm dof})=316.7/414$, with $\Delta\chi^2$ reductions of 50 and 30 with respect the {\tt Model 2}, respectively, for comparable degrees of freedom); however, it did provide parameters that describe a physically reliable geometry.
As reported in Table \ref{reflection_table}, we determined an inner disk radius of $22^{+25}_{-7} ,{\rm R{g}}$ and $44^{+77}_{-20} ,{\rm R{g}}$, consistent with both the results of {\tt Model 1} and the findings reported by \citet{Marino_2023MNRAS.525.2366M}. 
Most notably, we obtained a well-constrained measurement of the system inclination angle, with values of $31^{+8}_{-5}$ and $20^{+13}_{-20}$ degrees for the two datasets, respectively.
%An F-test comparing this model to one including the {\tt diskline} component yielded a probability of chance improvement of approximately $1.6 \times 10^{-12}$, implying that the inclusion of the reflection component significantly enhances the quality of the fit, with a confidence level exceeding 7$\sigma$.

\subsection{Other reflection models}
We also modeled the spectrum using two additional self-consistent reflection models: {\tt RelxillCp} \citep{Dauser_2016A&A...590A..76D} and {\tt Relxillns} \citep{Garcia_2022ApJ...926...13G}.
Both models are part of the {\tt Relxill} suite and employ an empirical broken power law to characterize the emissivity profile. The primary distinction between them lies in the nature of the incident spectrum: {\tt RelxillCp} uses the {\tt nthcomp} model \citep{Zdziarski_1996MNRAS.283..193Z} as the primary continuum with a fixed seed photon temperature of 0.05 keV, whereas {\tt Relxillns} adopts a blackbody spectrum with a temperature described by {\tt ${\rm kT_{bb}}$}.

Since both models assume an empirical broken power law for the emissivity, they include two distinct emissivity indices corresponding to regions above and below a specified break radius, ${\rm R_{br}}$, from the compact object. However, in our analysis, the data lack sufficient sensitivity to detect a significant difference between these two regions. Consequently, we opted to link the two emissivity indices, treating them as a single parameter.
As previously done with {\tt Model 3}, the relative reflection fraction was fixed at a negative value, specifically set to -1, while allowing the normalization to vary. For {\tt RelxillCp}, the Comptonization index gamma and the electron temperature were tied to those of {\tt comptb}, whereas for {\tt Relxillns}, the blackbody temperature, ${\rm kT_{bb}}$, was linked to the electron temperature of {\tt comptb}.
%In both fits, the values of the equivalent hydrogen column {\tt nH} were kept fixed to those obtained with the previous models, as preliminary fits showed that the best-fit values were the same, even though the error calculation for the abundance of Iron in the ISM only provided an upper limit of 0.8 for both models and in both fits. 

The model with {\tt Relxillns} (hereafter referred to as {\tt Model 4}) provided an excellent fit, yielding a $\chi^{2}$(d.o.f.) of 763(811) for the first fit set, and 326(414) for the second (although there was no significant improvement over {\tt Model 3} in either case). The best-fit parameters, including the inclination angle and inner radius are consistent with the previously obtained values. The rest of the parameters, reported in Table \ref{reflection_table}, will be discussed in more detail in the following sections.

Unlike {\tt Models 2} and {\tt 3A}, the {\tt RelxillCp} model ({\tt model 4}) was unable to reach stable values for the inner radius and inclination angle. As a result, we attempted to fix these parameters to the values obtained previously. However, the model failed to adequately describe the data. 
Although the $\chi^{2}$ value was statistically acceptable (850.02 with 810 d.o.f. and 489.66 with 431 d.o.f.), the data showed peculiar residuals in the energy range corresponding to the iron emission line.  Furthermore, the inclusion of the {\tt RelxillCp} component resulted in a worsening of the $\chi^{2}$.
%An F-test comparing {\tt model 4A} with {\tt Models 2} and {\tt 3A} showed a probability of chance improvement of $\sim 2 \times 10^{-24}$, confirming that the inclusion of the {\tt Relxillns} and {\tt Rfxconv} components enhances the fit quality with a confidence level greater than 7$\sigma$.
It is, therefore evident that the use of the {\tt RelxillCp} component does not provide an improved fit to the data. 
Consequently, all further discussion of the system will be based on the best-fit results obtained with {\tt Models 2} and {\tt 3A}.
The inadequacy of the model in representing the data is likely attributed to the fact that {\tt RelxillCp} includes a Comptonized spectrum with a seed photon temperature fixed at 0.05 keV. This value contrasts with the seed photon temperature determined for this source, which is approximately 0.8 keV and 1.0 keV for the two datasets, respectively. 

\begin{table*}[]    %quella nuova
\caption{\label{reflection_table} Best-fit parameters obtained applying {\tt Model 3 (Rfxconv)} and {\tt Model 4 (Relxillns)} to the data. Uncertainties are calculated at the 90\% confidence level.}
\centering
\renewcommand{\arraystretch}{1.3} % Aumenta l'interlinea
\begin{threeparttable}
\begin{tabular}{llcccc}
\hline
\multicolumn{1}{l}{\multirow{2}{*}{Model}} & \multirow{2}{*}{Parameter} & \multicolumn{2}{c}{{\tt Model 3}\tnote{\#}} & \multicolumn{2}{c}{\tt Model 4\tnote{\#}} \\ 
\multicolumn{1}{c}{}                       &                            & Oct          & 15 Apr          & Oct           & 15 Apr          \\ \hline
{\tt Tbfeo}                                & nH ($10^{22}$)             & $0.179^{+0.011}_{-0.009}$ & $0.173^{+0.011}_{-0.0079}$ & $0.181^{+0.006}_{-0.005}$ & $0.180^{+0.009}_{-0.007}$ \\
                                           & O                          & $1.35^{+0.06}_{-0.08}$    & $1.333^{+0.13}_{-0.098}$   & $1.36^{+0.07}_{-0.06}$    & $1.340^{+0.10}_{-0.01}$    \\
                                           & Fe                         & $0.5 \pm 0.4$             & $0.81^{+0.45}_{-0.5}$      & $0.5 \pm 0.3$             & $0.7 \pm 0.5$     \\
                                           
{\tt Expabs }                              & E (keV)                    & {\tt ${\rm kT_{s}}$} \tnote{$\dag$}   & {\tt ${\rm kT_{s}}$}\tnote{$\dag$}   & {\tt ${\rm kT_{s}}$}\tnote{$\dag$}   & {\tt ${\rm kT_{s}}$}\tnote{$\dag$}   \\
{\tt Powerlaw }                           & Index                      & $2.7 \pm 0.2$             & $2.77^{+0.18}_{-0.28}$     & $2.606$ \tnote{*}         & $2.8^{+0.2}_{-0.3}$     \\
                                           & Norm                       & $0.4^{+0.3}_{-0.2}$       & $0.46^{+0.22}_{-0.4}$      & $0.25 \pm 0.04$           & $0.7^{+0.6}_{-0.5}$    \\

{\tt diskbb }                              & ${\rm T_{in}}$ (keV)                  & $0.60^{+0.15}_{-0.05}$    & $0.684^{+0.083}_{-0.057}$  & $0.61 \pm 0.06 $          & $0.726^{+0.11}_{-0.08}$  \\
                                           & Norm                       & $910^{+600}_{-590}$       & $660^{+260}_{-210}$        & $960^{+330}_{-230}$       & $570^{+320}_{-240}$       \\

{\tt Comptb  }                             & ${\rm kT_{s}}$ (keV)                  & $0.76^{+0.14}_{-0.05}$    & $0.978^{+0.096}_{-0.069}$  & $0.79^{+0.10}_{-0.08}$    & $1.1^{+0.2}_{-0.1}$    \\
                                           & $\Gamma$                   & 3\tnote{*}               & 3\tnote{*}               & 3\tnote{*}               & 3\tnote{*}               \\
                                           & $\alpha$       & $0.85^{+0.03}_{-0.02}$   & $0.877^{+0.044}_{-0.038}$  & $0.93^{+0.04}_{-0.03}$    & $1.01^{+0.12}_{-0.06}$  \\
                                           & $\Delta$                   & 0\tnote{*}               & 0\tnote{*}               & 0\tnote{*}               & 0\tnote{*}               \\
                                           & ${\rm kT_{e}}$ (keV)                  & $2.88 \pm 0.02$          & $2.834^{+0.029}_{-0.033}$  & $2.91 \pm 0.04$           & $2.86^{+0.06}_{-0.04}$  \\
                                           & log (A)                    & 8\tnote{*}               & 8\tnote{*}               & 8\tnote{*}               & 8\tnote{*}               \\
                                           & Norm                       & $0.051 \pm 0.001$        & $0.065 \pm 0.002$ & $0.052 \pm 0.001$        & $0.067 \pm 0.002$ \\

{\tt Reflection}                           & Index                      & $-2.6^{+0.7}_{-1.2}$    & $-1.7^{+0.8}_{-0.5}$     & $-2.8^{+1.5}_{-0.7}$    & $2.3$\tnote{*}           \\
                                           & ${\rm R_{in}}$ (R${\rm g}$)           & $22^{+25}_{-7}$           & $10 \pm 2\tnote{$\odot$}$                          & $<35$                    & $10^{+3}_{-2}\tnote{$\odot$}$                       \\
                                           & ${\rm R_{out}}$                       & $1.0$\tnote{*}          & $1.0$\tnote{*}          & $1.0$\tnote{*}          & $1.0$\tnote{*}          \\
                                           & Incl (deg)                 & $31^{+9}_{-5}$       & $15^{+16}_{-13}$          & $34^{+7}_{-5}$      & $35^{+6}_{-4}$    \\
                                           & rel refl               & $-0.157^{+0.053}_{-0.032}$ & $-0.085^{+0.037}_{-0.072}$ & -1\tnote{*}              & -1\tnote{*}        \\
                                           & Fe$_{abund}$               & $0.28^{+0.2}_{-0.1}$     & $0.5^{+0.4}_{-0.2}$    & $<0.77\tnote{$\bigstar$}$         & $<1.26\tnote{$\bigstar$}$          \\
                                           & log N                      & -                          & -                    & <17.6      & <16.5  \\ 
                                           & log$\xi$                      & $2.69^{+0.05}_{-0.19}$ & $2.8^{+0.1}_{-0.2}$     & $3.1^{+0.2}_{-0.3}$     & $2.8 \pm 0.2$   \\ 
                                           & Norm (10$^{-4}$)      & - & -  & $9^{+3}_{-2}$  & $13^{+4}_{-3}$  \\ 
                                           \hline
                                           & $\chi^2$/dof               & $762/810$             & $317/414$              & $763/810$              & $326/414$             \\ \hline
\end{tabular}
\begin{tablenotes}
\item[$\dag$] Linked to the seed-photon  temperature.
\item[*] Kept frozen during the fit.
\item[$\odot$] Linked to the emitting radius of the {\tt diskbb} component.
\item[$\bigstar$] Upper limit at 90\% confidence level obtained fixing the seed photons and blackbody temperature.
\item[\#]  {\tt Model 3: const*TBfeo*(expabs*powlaw+rdblur*rfxconv*Comptb+Comptb+diskbb)} 
 {\tt Model 4: const*TBfeo*(expabs*powlaw+Relxillns+Comptb+diskbb)}
\end{tablenotes}

\end{threeparttable}
\end{table*}

\subsection{Polarization analysis}

\begin{figure}
    \centering
    \includegraphics[width=0.475\textwidth]{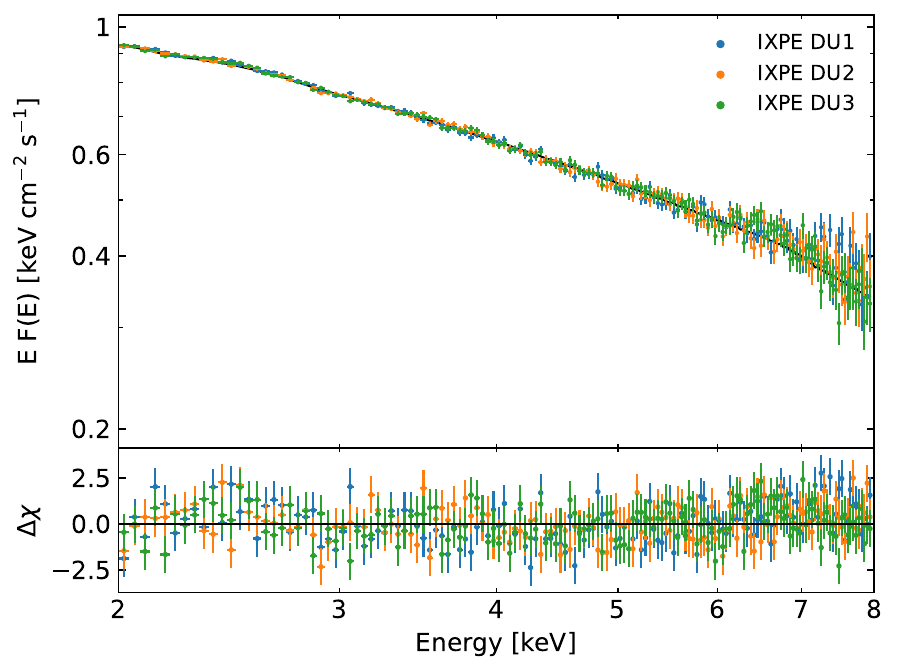}
    \includegraphics[width=0.475\textwidth]{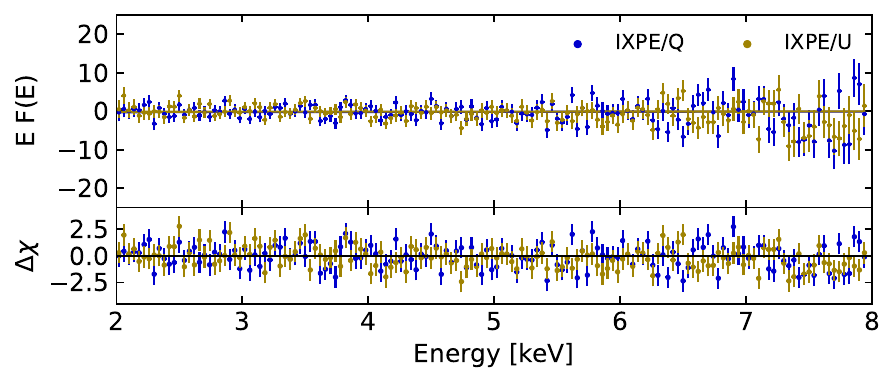}
    \caption{Deconvolved \ixpe $I$ spectra of each DU (top panels), $Q$ and $U$ Stokes spectra (bottom panels), with the best-fit model (October+April 16, Table \ref{reflection_table}) and residuals in units of $\sigma$.}
    \label{fig:IXPE.Spectra}
\end{figure}

To study the polarization of 4U 1820--30, we first applied the \texttt{polconst} model to the whole best-fit model obtained for each dataset. We started by fitting with XSPEC the $I$, $Q$, and $U$ spectra averaged throughout the entire IXPE observation (Fig. \ref{fig:IXPE.Spectra}). In Table \ref{tab:Pol}, the polarization degree (PD) and polarization angle (PA) measured in different energy ranges are reported. The results are the same for all the best-fit models considered and are consistent within errors with that obtained by \cite{DiMarco.4U1820} with \textsc{ixpeobssim}: the PD increases from only an upper limit of 1.2\% (at the 99\% confidence level)  below 4 keV up to $8.0\% \pm 3.6\%$ (at the 90\% confidence level) in the 7--8 keV energy bin, but remained unconstrained in the 6--7 keV energy range where the (unpolarized) Fe line is observed. Polarization in the 7--8 keV energy range is detected at 3$\sigma$.  
{Since the observed sudden increase in polarization in the 7–8 keV band} is peculiar and has not been observed in any other source, we performed a new spectropolarimetric analysis using the best-fit models obtained from the NICER and \textit{NuSTAR} datasets. In particular, differently from \cite{DiMarco.4U1820}, we were able to fit the data including the reflection component, which can affect the polarization signal even if its flux is relatively small.

We performed the spectropolarimetric analysis by applying a constant polarization to each spectral component using \texttt{polconst} and fixing all spectral parameters to the best-fit values obtained with NICER and \textit{NuSTAR}. Due to the limited IXPE band-pass, it is challenging to constrain both the PD and PA for all spectral components together. Specifically, the Comptonized and reflected photons tend to overlap significantly and share a similar spectral shape; therefore, some assumptions are needed to estimate the polarization properties. In particular, for all the cases, the polarization of the power law was tied to that of {\tt Comptb} since it should originate from hybrid thermal or non-thermal Comptonization (Sect. \ref{sec:Continuum.Spectrum}), while we assumed the same polarization angle for reflection and Comptonization (see e.g., \citealt{Ursini.etAl.2023,Gnarini.etAl.2024}). We first tried to divide the IXPE observation into two segments simultaneously with NICER and \textit{NuSTAR} for each period (i.e., October+April 16 and April 15), but the statistic was not enough to estimate the polarization of the different spectral components. During the spectropolarimetric analysis, we considered only the best fit of Model 3 obtained using the October+April 16 dataset (Table \ref{reflection_table}). {No matter whether we separate the observation into two segments or use the different models in Table \ref{reflection_table}, we would expect to obtain similar results for the polarization. This is not only because the relative fluxes of each component with respect to the total flux do not differ significantly, but also because the various observations are broadly consistent, with the only notable difference being that the April 15 spectrum is slightly harder.}
%Whether separating the observation into two segments or using the different models in Table \ref{reflection_table}, the relative fluxes of each component with respect to the total flux are not very different; therefore, we expect similar results for the polarization.

Table \ref{tab:Pol.Comp} reports all the results obtained from the spectropolarimetric analysis using \texttt{polconst} for each component. As reported in \cite{DiMarco.4U1820} we started by assuming only one polarized component with the others having null polarization: for both best-fit models of the October+April 16 dataset, only upper limits were obtained for the Comptonization and the disk components, while the reflection seems to be very highly polarized as expected (up to $\approx 10\%$; see also \citealt{Matt.1993}). Then, we tried to consider all components polarized. We first fixed the PD of Comptonization at 1\%, which is a reasonable value for typical spreading layer configurations \citep{Ursini.etAl.2023,Farinelli.etAl.2024,Bobrikova.etAl.2024}, leaving the PD of the disk and reflection free to vary; in this case, the fit improves ($\chi^2$/d.o.f. = 1395.8/1335), but both components end up more polarized than expected, with the reflection reaching $\approx 16\%$ and the disk at about 9\%; this result is significantly higher than the classical results for an electron scattering dominated atmosphere on the accretion disk observed at $i \approx 35$\degr \citep{Chandrasekhar.1960}. These high values may be due to the relatively small flux of these components, which also leads to large errors. On the other hand, we tried to fix the PD of the reflected photons at 10\% \citep{Matt.1993}, leaving free to vary the polarization of the disk and the Comptonization: the fit is better than the cases with only one polarized component ($\chi^2$/d.o.f. = 1396.1/1335), the disk appears to be again much more polarized than expected ($\approx 8\%$), while the PD for Comptonization is consistent with the expected polarization for spreading layer geometries ($\approx 1\%$). For both scenarios, the disk also seems to be polarized orthogonally to the Comptonization, which is the main contribution to the polarized signal in the 2--8 keV band. 

\section{Discussion}
We analyzed the average broad-band spectra of three simultaneous NICER and \textit{NuSTAR} observations of 4U 1820-30, focusing on the possible presence of reflection features. As seen in the previous section, we confirmed the presence of a intense reflection component, whose inclusion (implemented in {\tt Models 2} and {\tt 3A}) significantly improved the fit, while flattening the residuals and providing valuable insights into the system parameters, namely, describing a geometry that is consistent with current literature results and is physically reliable for this system. 

\subsection{Continuum spectrum}\label{sec:Continuum.Spectrum}
The continuum emission is aptly described by a disk blackbody component plus a Comptonized and a power-law component. 
All future considerations regarding the continuum will be based on the parameters obtained with {\tt Model 3}, as the application of {\tt model 4} yields the same results.
The temperature of the blackbody component does not show a significant variation between the two fits, at \(0.60^{+0.15}_{-0.05}\,\mathrm{keV}\) during the October and April 16 observations and \(0.68^{+0.08}_{-0.06}\,\mathrm{keV}\) on the April 15 one. 
%What varies, however, is the normalization, which directly impacts the inferred size of the emission region.
The normalization also seems to be consistent within the errors, as we would expect the size of the emission region to be.
The emission region size can be derived using the formula:
\[
K_{\mathrm{disk}} = \left(\frac{R_{\mathrm{disk}}}{D_{10 \,\mathrm{kpc}}} \right)^2 \cos{\theta},
\]
where \(K_{\mathrm{disk}}\) is the normalization of the blackbody component, \(\theta\) is the inclination angle of the disk with respect to the line of sight, and \(D_{10}\) is the distance to the source in units of 10 kpc.
Assuming the latest estimated distance for the source of \(8.0 \pm 0.3\,\mathrm{kpc}\) \citep{Baumgardt_2021MNRAS.505.5957B} and using the obtained values for the normalization and inclination angle, we derived apparent blackbody radii, \(R_{\mathrm{bb}}\), of \(26^{+8}_{-10} \,\mathrm{km}\) and \(21^{+4}_{-3} \,\mathrm{km}\) for the two datasets, respectively.
Correcting these values by a hardening factor of \(f = 1.7\), as suggested by \citet{Shimura_1995ApJ...445..780S}, which accounts for the effects of electron scattering in the atmosphere of the accretion disk, we obtained radii of \(R_{\mathrm{eff}} = 44^{+13}_{-18} \,\mathrm{km}\) and \(36^{+7}_{-6} \, \mathrm{km}\), respectively. 
These results, in addition to being perfectly consistent with the inner disk radius obtained from the reflection component, also indicate that the disk emission originates from the same region in all three observations: October 11-12, April 16, and April 15.

As previously discussed, the power-law component is introduced to account for excess emission at high energies that cannot be explained by simple thermal Comptonization models alone.
These components are often interpreted as originating from non-thermal processes, or hybrid thermal or non-thermal Comptonization, which involves both thermal and non-thermal electron populations in the plasma \citep[e.g.,][]{dai_2007ApJ...667..411D}.
Over the years, models addressing a self-consistent description of this emission have been developed, such as {\tt EQPAIR} by \cite{Coppi_1999ASPC..161..375C}. 
In the case of 4U 1820-30, the presence of a hard tail with a photon index of approximately 2.4 -- consistent with our findings -- has previously been observed at high energies (around 50 keV) by \citet{Tarana2007ESASP.622..433T}. The authors suggest that its origin could be attributed either to non-thermal electrons or to thermal emission from plasmas with a relatively high temperature.
In our analysis, while the inclusion of a power-law component is statistically significant, improving the $\chi^2$ with a confidence level higher than 7$\sigma$, its flux contribution becomes relevant only at higher energies (approximately $30\%$ of the total flux between 30 and 35 keV). Furthermore, its presence or absence does not impact the description of the spectrum at lower energies. For this reason, we do not delve into the use of other self-consistent models to explore its physical origin.
\begin{figure*}
    \centering
    \includegraphics[width=0.45\linewidth]{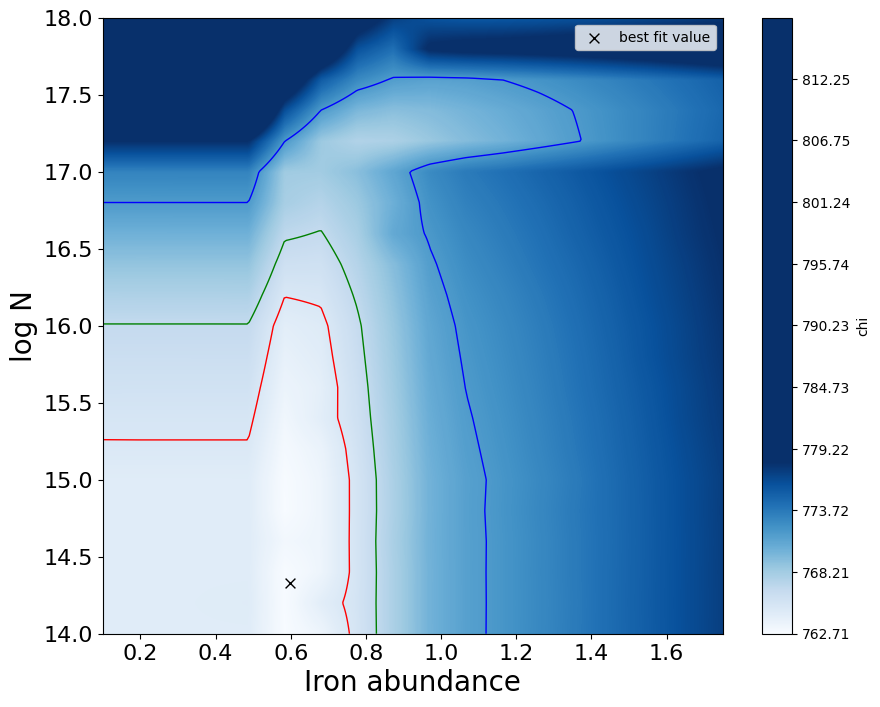} \quad
    \includegraphics[width=0.45\linewidth]{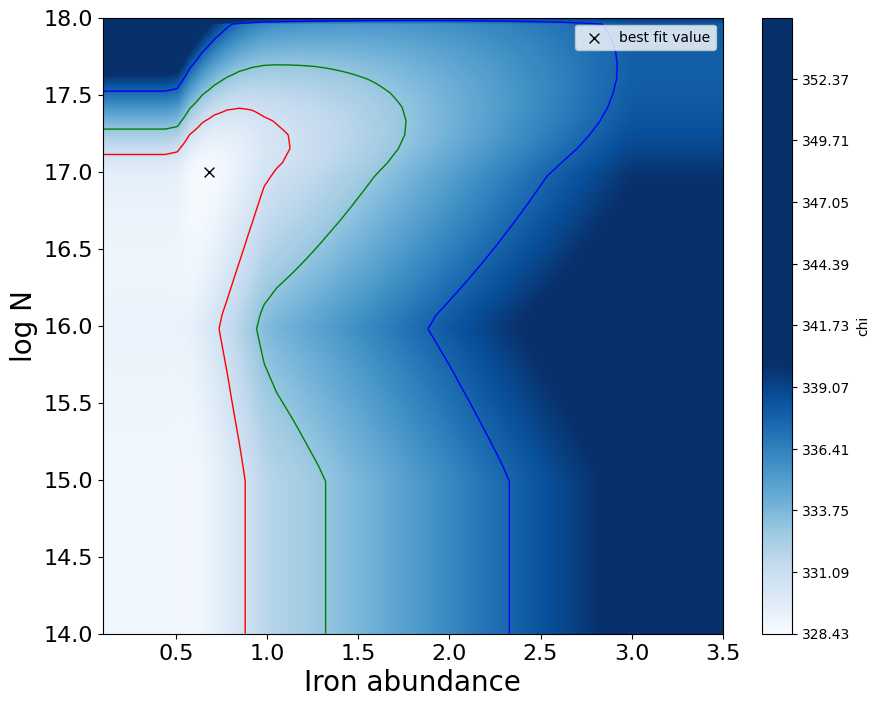}
    \caption{\label{steppar_iron} Contour plots for the logarithmic electron density (log N) and the iron abundance of the disk obtained applying {\tt model 4} to October and April 16 data (left panel) and April 15 data (right panel). The contours represent the 1$\sigma$, 2$\sigma$, and 3$\sigma$ confidence levels, and the cross marks the best-fit values obtained from the best fit. {The axis limits have been rescaled for visual purposes}.}
    \label{fig:enter-label}
\end{figure*}

The main factor preventing us from simultaneously analyzing all three datasets (thereby isolating the April 15 observation and resulting in its divergent spectral shape) seems to be the Comptonization component. Indeed, while the electron temperature remains consistent across the two fits ($2.88 \pm 0.02$ keV and $2.83^{+0.04}_{-0.03}$ keV), the seed photon temperature (${\rm kT}_{{\rm s}}$) shows a variation, increasing from $0.76^{+0.14}_{-0.05}$ keV to $0.98^{+0.09}_{-0.04}$ keV, resulting in a spectral shift towards higher energies. 
Such a difference in the seed-photon temperature may indicate a slightly distinct emission region for the seed photons, as the seed photon radius,  $R_{{\rm sp}}$, can be expressed as \citep[e.g.,]{Anitra_2021A&A...654A.160A}: 
\begin{equation}
R_{{\rm sp}} = 3 \times 10^4 \, f_{{\rm seed}}^{1/2} \, {\rm kT}_{{\rm s}}^{-2} \, {\rm D} \,  \,{\rm (km)},
\end{equation}
where  D is the source distance and $f_{{\rm seed}}$ represents the flux of the seed photon spectrum before being Comptonized in the corona.

If we consider that when log A assumes its minimum value (-8), the \texttt{comptb} model is reduced to the standard \texttt{bb} (blackbody) model in XSPEC \citep{Farinelli_2008ApJ...680..602F}, the seed photon flux can be determined by setting log A =-8 and computing the flux for the Comptonization-only model. 
Thus, we can obtain a seed photon flux of \((4.3 \pm 0.4) \times 10^{-9} \, \text{ergs/cm}^2/\text{s}\) for the October and April 16 datasets, and \((5.5 \pm 0.5) \times 10^{-9} \, \text{ergs/cm}^2/\text{s}\) for the April 15 one.  

These flux values correspond to seed photon radii of \(27^{+10}_{-4}\) km and \(19^{+4}_{-3}\) km for the two fits, suggesting that the emission regions are consistent (within the uncertainties) across all three observations. 
Such high values also exclude the NS surface as the source of the seed photons, pointing instead to a more extended region, which is likely to be the boundary layer connecting the inner edge of the accretion disk to the NS surface.

Previous studies have highlighted variations in the seed photon emission region during the spectral evolution of the source. 
As discussed, 4U 1820-30 exhibits intrinsic luminosity modulation, fluctuating between a low mode (\(L_{\text{low}} \sim 8 \times 10^{36} \, \text{erg s}^{-1}\)) and a high mode (\(L_{\text{high}} \sim 6 \times 10^{37} \, \text{erg s}^{-1}\)) over a timescale of approximately 170 days.
Analyzing RXTE data, \citet{Titarchuk_2013ApJ...767..160T} reported an increase in electron temperature from 2.9 to 21 keV with a corresponding decrease in the normalization of the Comptonization component. This implies a reduction in the size of the seed-photon region when transitioning from the high mode to the low mode.
More recently, \citet{Marino_2023MNRAS.525.2366M} confirmed this trend,  through an extensive X-ray campaign involving NICER, \textit{NuSTAR}, and AstroSat observations. Their analysis showed that in the high mode, seed photons are emitted from a larger region, roughly 15 km, while in the low mode, the emission region shrinks to about one-third of this size.
The authors attributed these variations to changes in the mass accretion rate: during the high mode, a higher accretion rate feeds a larger boundary layer, producing more photons that cool the corona. Conversely, in the low mode, the reduced accretion rate results in a smaller seed photon emission region and fewer photons available to cool the corona.

Our results describe a Comptonization spectrum with an electron temperature of \(\sim 2.8 \, \mathrm{keV}\), where the seed photons are emitted from a boundary layer %%at a distance of 
of radius \(\sim 23 \, \mathrm{km}\). %%from the NS. Moreover, the luminosity calculated across all models is nearly constant at approximately \((8-9) \times 10^{37} \, \mathrm{erg \, s^{-1}}\) (in the \(0.1-100 \, \mathrm{keV}\) energy range). These findings strongly support the geometry proposed by \citet{Marino_2023MNRAS.525.2366M}, placing the source in an extremely high state.
{The larger emission region we infer in this work, compared to \cite{Marino_2023MNRAS.525.2366M}, likely reflects a difference in accretion state. While the authors of the cited study reported fluxes on the order of $1 \times 10^{-9} {\rm erg \, cm^{-2} \, s^{-1}}$, we measured significantly higher unabsorbed fluxes, of around $9.2 \times 10^{-9} {\rm erg \, cm^{-2} \, s^{-1}}$, corresponding to a luminosity of approximately $7 \times 10^{37} {\rm erg \, s^{-1}}$, suggesting that the source was in a very high mode during our observations. This supports the idea that variations in the seed photon region are driven by changes in mass accretion rate, as  proposed  in other literature works.}

\subsection{Reflection component}
The relativistic Fe K line, potentially linked to a reflection component, has long been a subject of debate in this source, with multiple studies in the literature have contradicted each other over time.
Early observations with Ginga \citep{Sansom1989PASJ...41..591S}  and RXTE \citep{Bloser2000ApJ...542.1000B} found no evidence of an Fe K emission line. It was only in 2004, with the work of \cite{Ballantyne2004ApJ...602L.105B}, that a 6.4 keV emission line was detected during a super-burst observed with RXTE.
Subsequently, \cite{Cackett_2008ApJ...674..415C} used \textit{Suzaku} data and \cite{Titarchuk_2013ApJ...767..160T}  used \textit{Beppo}SAX spectra in their studies, while, more recently, \cite{Marino_2023MNRAS.525.2366M} provided evidence for a broad, asymmetric Fe K emission  described using the {\tt diskline} and {\tt Laor} models. 
However, contrasting results have been reported by \cite{Costantini2012A&A...539A..32C} and \cite{Mondal_2016MNRAS.461.1917M}, where the \textit{NuSTAR} and \textit{Swift} data did not detect any significant Fe K feature.

 Within this scenario, our analysis clearly shows the presence of a strong reflection component in the spectrum, with a significant percentage of flux contribution (about 7\% of the total flux for all datasets). 
The reflection component was described using two different self-consistent models, both of which improved the chi-squared value with a confidence level exceeding 7 sigma compared to the model with only the {\tt diskline}, proving the presence of a reflection component in the source spectral emission, at least during the observations considered here.

%The best-fit parameters obtained from the two models align well with expectations. 
The inner radius indicates emission from a region close to the NS, with ${\rm R_{in}} = 22^{+25}_{-7}\, {\rm R_{g}}$ for the fit obtained using {\tt Model 3} and an upper limit of 35 ${\rm R_{g}}$ for the fit with {\tt model 4}.
These radii are consistent with the blackbody radius obtained from our fits, confirming that the disl is placed just outside the boundary layer, as in the geometry proposed by \cite{Marino_2023MNRAS.525.2366M}.
However, for the April 15 dataset, the values of ${\rm R_{in}}$ are not well constrained. Specifically, with {\tt Model 3}, we obtained ${\rm R_{in}} = 44^{+77}_{-20}\, {\rm R_{g}}$, while the fit with {\tt model 4} did not yield a stable value for ${\rm R_{in}}$. Consequently, we opted to link it to the inner radius derived from the normalization of the {\tt diskbb} component.
Similarly, the error calculation for the emissivity index in {\tt Model 4} for the April 15 fits did not yield a stable value. Consequently, we fixed it to the best value obtained from the fit.
For the other fits, we derived an emissivity index of $-2.6^{+0.7}_{-1.2}$ and $-2.8^{+1.5}_{-0.7}$ for  October and April 16, respectively, and $-1.7^{+0.8}_{-0.5}$ for  April 15. {These results are consistent with values expected for illumination by an extended corona or a boundary layer close to the NS surface, rather than from a compact point source \citep{Dauser_2013MNRAS.430.1694D,Cackett_2010ApJ...720..205C}.}
%The emissivity index is expected to range between 2 and 3, depending on whether the reflection spectrum is primarily influenced by the illuminating flux from the central source (where the flux decreases with distance as $r^{-2}$) or by intrinsic disc emission (which decreases approximately as $r^{-3}$).

The most significant result of our analysis is a consistent measurement of the system inclination angle using both models and across all observations: $31^{+9}_{-5}$ and $34^{+7}_{-5}$ degrees for the October and April 16 data, and $15^{+16}_{-13}$ and $35^{+6}_{-4}$ degrees for the April 15 data. 
These results match with the low inclination hypothesis reported in the literature \citep{Anderson_1997ApJ...482L..69A}.
%Regarding the ionization parameter $\xi$ (defined as $\xi = 4\pi F_{x} /n_{e}$, where ${\rm F_{x}}$ is the X-ray flux and ${\rm n_{e}}$ is the electron number density \citep{Fabian_2000}), we observed a tendency for the model incorporating {\tt relxillns} to yield higher values for this parameter. Specifically, for the \apr fit, both models provided comparable values of  and . However, in the \oct fit, Model 2 returned a  of , while the inclusion of {\tt relxillns} significantly increased this value to TATA.

\subsection{Iron abundance}
During the fit, we allowed for variations in the iron abundance in the system and in ISM, as this adjustment improved the fit, especially when the abundance was set lower than the solar one.
Application of \textsc{Model 3} provides iron abundances in the ISM of $0.5 \pm 0.4$ and $0.8^{+0.4}_{-0.5}$, and subsolar iron abundances in the disk of $0.5^{+0.4}_{-0.2}$ and $0.3^{+0.2}_{-0.1}$, for the October plus the April 16 and April datasets, respectively.
For \textsc{model 4}, we obtained ISM iron abundances of $0.5 \pm 0.3$ and $0.7^{+0.4}_{-0.5}$, in line with the ones obtained through \textsc{Model 3}, but the abundance of iron in the disk remained undefined. 
To determine a better constrain on the abundances, we fixed the seed photon temperature for the Comptonization component, which is a parameter that was already well-constrained in previous fits. This resulted in an upper limit of 0.77 for the October and April 16 data, with 1.26 for April 15.

Certain parameters such as the iron abundance are highly sensitive to external factors, including the electron density in the disk. As shown by \citet{Ding_2024ApJ...974..280D}, in reflection models with low electron densities ($\log N_e \lesssim 15 $), the inferred iron abundance often exceeds the solar value. This effect arises because the strength of iron lines is overestimated to compensate for the limitations of atomic physics in low-density regimes.
At higher electron densities ($\log N_e \gtrsim 15 $), recombination rates are significantly reduced, leading to higher gas ionization. This results in stronger emission lines, such as those of iron and oxygen, while the continuum below 2 keV becomes less prominent. %Under these conditions \textcolor{red}{QUALI??}, the inferred iron abundances become more reliable and are often consistent with solar values.

To investigate whether the underabundance of iron we detected is biased by an electron density value, we employed the \textsc{steppar} command in XSPEC. This means we performed fits by varying the values of the two parameters within a given range (from 0.1 to 3 for the iron abundance and from 14 to 20 for the electron density). This approach allowed us to evaluate all possible combinations of $\log N_e$ and iron abundance, measuring how the $\chi^2$ value changes relative to the best fit for each combination. By doing so, we identified parameter combinations that might result in an equivalent spectral fit. 
The contour plots in Fig. \ref{steppar_iron} show the parameter space explored using {model 4} for both datasets. Even if the regions do not clearly define a lower limit for the parameters, they reveal how (at least within a 1$\sigma$ confidence level) the lowest chi-square values consistently correspond to iron abundances that are always below  solar.

The subsolar iron abundance observed in both the disk material and the interstellar medium can likely be explained by the source location within the globular cluster NGC 6624. Globular clusters are composed of old population II stars, which exhibit chemical compositions that differ significantly from solar values. These clusters are particularly notable for their variability in the abundances of heavy elements such as iron, likely due to their distinct evolutionary histories \citep{Gratton_2012A&ARv..20...50G}.
The iron abundance value we derived is consistent with the findings of \citep{Heasley_2000AJ....120..879H}. The authors used high-resolution infrared spectroscopy with the Hubble Space Telescope to measure an iron abundance for NGC 6624 of $\rm{[Fe/H] = -0.63 \pm 0.1}$, where [Fe/H] is the logarithmic ratio of the stellar metal abundance relative to solar values.
Furthermore, the presence of a  white dwarf in the system \citep{Stella_1987ApJ...315L..49S,Rappaport_1987ApJ...322..842R}, rather than a main-sequence star, might also play a role in altering the metallicity within the disk. It is indeed suggested by \cite{Koliopanos2021MNRAS.501..548K} that the chemical composition of the accreted material in UCXBs can differ significantly from that of standard X-ray binaries. In particular, a hydrogen-deficient composition, resulting from the presence of a white dwarf donor, could alter the amounts of carbon, oxygen, and iron in the disk, making the Fe K line either more or less difficult to observe.

\subsection{Energy dependence of X-ray polarization}
From the relative photon fluxes of each component, we can notice how the polarization significantly increases where the contribution of the disk drops: although the X-ray emission is always dominated by Comptonization in the IXPE band, at low energies, where no detection is found, the disk represents about $20-25\%$ of the total flux; while it is only 0.1\% of the total flux in the 7--8 keV, where the PD is higher (Table \ref{tab:Pol}). This is in line with a disk polarized orthogonally to both Comptonization and reflection, as we obtained from spectropolarimetric analysis (Table \ref{tab:Pol.Comp}). Since the Comptonization is the main source of the observed polarization signal, its PD can not be very high over the entire 2--8 keV energy band; otherwise we would have observed significant polarization even at low energies. 

{\cite{Nitindala_2025A&A...694A.230N} recently showed that high levels of polarization can be achieved through scattering in accretion disk winds. However, their results indicate that at low inclinations (e.g., in the case of 4U 1820–30), the contribution of disk winds to the polarization signal is strongly suppressed, due to the geometry of the system and the limited scattering along the line of sight. Furthermore, strong disk winds generally produce clear spectral signatures, such as absorption lines or P-Cygni profiles, which have never been observed in the X-ray spectra of this source. While a minor contribution from a weak or highly ionized wind cannot be entirely excluded, it is unlikely that such a component alone accounts for the polarization trend we have observed.}

Therefore, we tried to study the behavior of the polarization as a function of energy using the \texttt{pollin} and \texttt{polpow} models. These two models describe a polarization with a linear or a power-law dependence on energy, respectively. In particular, since we do not find any significant rotation in the PA with energy for this source, the PA is considered constant, while the PD can vary with energy $E$, respectively, as:
\begin{align}
    & {\tt pollin}:  \text{PD}(E) = \text{PD}_\text{1\,keV} + (E-1\,\mathrm{keV}) \times A_\text{PD} ~, \\
    & {\tt polpow}: \text{PD}(E) = \text{PD}_\text{1\,keV} \times E^{\alpha_\text{PD}} ~.
\end{align}
We applied these models to both the Comptonization and the reflection components: in fact, it is possible that at least one of these two should vary with energy to explain the behavior of the polarization at higher energies. We report the results obtained using \texttt{pollin} or \texttt{polpow} in Tables \ref{tab:Pollin.Comp} and \ref{tab:Polpow.Comp}. We considered two different scenarios: one with constant PD of the reflection (fixed at 10\%; \citealt{Matt.1993}) and the PD of the Comptonization variable with energy \citep{Gnarini.etAl.2022,Bobrikova.etAl.2024}; and the second,  with a constant PD of the Comptonization (fixed at 1\%; \citealt{Ursini.etAl.2023,Gnarini.etAl.2024,Farinelli.etAl.2024,Bobrikova.etAl.2024}) and that of the reflection variable with energy \citep{Podgorny.etAl.2022,Podgorny.etAl.2023}. These cases provide slightly better fits to the data with respect to using only \texttt{polconst}, but not so much better as to have a strong claim of an increasing trend of PD with energy. In any case, none of these results are able to reproduce the high polarization in the 7--8 keV bin, in particular, given the relatively low inclination of the system derived by the spectral analysis: when the PD of Comptonization increases with energy, the maximum PD obtained at high energies is $\approx 3.5\%$, which is not consistent with the observed PD, considering the 90\% confidence level error; when the PD of the reflection increases with energy, it will reach extremely high values in the 7--8 keV range, $\approx 30\%$ considering {\tt pollin}, and $\approx 40\%$ considering {\tt polpow}.

\begin{table*}[ht]
\centering
\renewcommand{\arraystretch}{1.2} % Adjust row height
\setlength{\tabcolsep}{8pt} % Adjust column separation
\caption{Polarization measured with XSPEC and percentage of flux of each component related to {\tt Model 3}.}
\label{tab:Pol}
\begin{tabular}{c|cc|ccc c}
\hline
\multicolumn{1}{l|}{\multirow{2}{*}{Energy range}} & {\multirow{2}{*}{PD}} & {\multirow{2}{*}{PA}}       &\multicolumn{4}{c}{Flux}       \\
\multicolumn{1}{l|}{}     &          &          &  Disk & Comp. & PL &  Refl. \\ \hline

2--8 keV & $<0.9\%$ & Unconstrained & 13.8\% & 77.1\% & 4.7\% & 4.4\% \\ 
2--4 keV & $<1.2\%$ & Unconstrained &  21.0\% & 69.4\% & 5.6\% & 4.0\% \\ 
4--6 keV & $1.7\% \pm 0.8\%$ & 121\degr $\pm$ 14\degr & 2.9\%  & 89.4\% & 3.6\% & 4.1\% \\ 
6--7 keV &  $<4.1\%$ & Unconstrained & 0.4\%  & 90.2\% & 2.9\% & 6.6\% \\ 
7--8 keV & $8.0\% \pm 3.6\%$ & 115\degr $\pm$ 14\degr & 0.1\%  & 91.5\% & 2.7\% & 5.7\% \\ \hline
\end{tabular}
\tablefoot{\centering The errors are at the 90\% confidence level, while the upper limits are reported at the 99\% confidence level.}
\end{table*}

\begin{table}
\caption{Polarization degree and angle for different scenarios, applying {\tt polconst} to each component.} 
\label{tab:Pol.Comp}      
\centering
\renewcommand{\arraystretch}{1.2}
\setlength{\tabcolsep}{8pt}
\begin{tabular}{lcc}
\hline
 Component & PD (\%) & PA (deg) \\   
\hline   
Disk & $<5.4$ & Unconstrained \\
Compton+PL & [0] & - \\
Reflection & [0] & - \\
\hline
Disk & [0] & - \\
Compton+PL & $<1.2$ & Unconstrained \\
Reflection & [0] & - \\
\hline
Disk & [0] & - \\
Compton+PL & [0] & - \\
Reflection & 8.9 $\pm$ 6.8 & 107 $\pm$ 33 \\
\hline
Disk & 8.7 $\pm$ 4.1 & 34 $\pm$ 14 \\
Compton+PL & [1] & 120 $\pm$ 13 \\
Reflection & 16.4 $\pm$ 13.2 & = PA Compton+PL \\
\hline
Disk & 7.6 $\pm$ 3.6 & 34 $\pm$ 15 \\
Compton+PL & 1.1 $\pm$ 0.7 & 120 $\pm$ 14 \\
Reflection & [10] & = PA Compton+PL \\
\hline
\end{tabular}
\tablefoot{
The errors are at the 90\% confidence level, while upper limits are reported at the 99\% confidence level. Parameters in square brackets are frozen. When only one component is polarized, the three best fits are statistically equivalent ($\chi^2$/d.o.f. = 1407.9/1337), while the fit improves in the last two cases ($\chi^2$/d.o.f. = 1396.1/1335).
}
\end{table}

\begin{table}
\caption{Polarization degree and angle for different scenarios, applying {\tt polconst} to the disk and {\tt pollin} to the other components.} 
\label{tab:Pollin.Comp}      
\centering
\renewcommand{\arraystretch}{1.2}
\setlength{\tabcolsep}{8pt}
\begin{tabular}{lccc}
\hline
Component & PD$_{1 \rm keV}$ (\%) & $A_{\rm PD}$ & PA (deg) \\   
\hline   
Disk & 5.9 $\pm$ 2.7 & - & 36 $\pm$ 15 \\
Comp.+PL & [0] & 0.3 $\pm$ 0.2 & 120 $\pm$ 12 \\
Reflection & [10] & - & = PA Comp.+PL \\
\hline
Disk & 7.0 $\pm$ 2.7 & - & 35 $\pm$ 14 \\
Comp.+PL & [1] & - & 120 $\pm$ 12 \\
Reflection & [0] & 4.4 $\pm$ 3.5 & = PA Comp.+PL \\
\hline
\end{tabular}
\tablefoot{
The errors are at the 90\% confidence level. Parameters in square brackets are frozen. The fits are statistically equivalent ($\chi^2$/d.o.f. = 1394.3/1335).
}
\end{table}

\begin{table}
\caption{Polarization degree and angle for different scenarios, applying {\tt polconst} to the disk and {\tt polpow} to the other components.} 
\label{tab:Polpow.Comp}      
\centering
\renewcommand{\arraystretch}{1.2}
\setlength{\tabcolsep}{8pt}
\begin{tabular}{lccc}
\hline
Component & PD$_{1 \rm keV}$ (\%) & $\alpha_{\rm PD}$ & PA (deg) \\   
\hline   
Disk & 5.9 $\pm$ 2.5 & - & 36 $\pm$ 15 \\
Comp.+PL & [0.1] & 1.7 $\pm$ 0.3 & 120 $\pm$ 12 \\
Reflection & [10] & - & = PA Comp.+PL \\
\hline
Disk & 6.8 $\pm$ 2.4 & - & 35 $\pm$ 13 \\
Comp.+PL & [1] & - & 120 $\pm$ 11 \\
Reflection & [1] & 1.8 $\pm$ 0.3 & = PA Comp.+PL \\
\hline
\end{tabular}
\tablefoot{
The errors are at the 90\% confidence level. Parameters in square brackets are frozen. The fits are statistically equivalent ($\chi^2$/d.o.f. = 1393.6/1335).
}
\end{table}

\section{Conclusions}
We present a detailed spectral-polarimetric analysis of the ultracompact X-ray binary 4U 1820-30 using data collected by NICER, \textit{NuSTAR}, and IXPE. 
The continuum emission is well described by a disk blackbody, a Comptonization component, and a power law. The inferred blackbody radius is consistent with the inner accretion disk, while the Comptonized spectrum reveals a stable electron temperature of $\sim$ 2.8 keV and seed photons originating from a boundary layer at $\sim$ 23 km from the NS. These findings strongly support the geometry driven by accretion rate variations proposed in \cite{Marino_2023MNRAS.525.2366M}, indicating that the source is in an extremely high state during the observations.

For the first time, we detected a strong reflection component in the spectrum, modeled using two different self-consistent reflection models. This allowed us to infer the a measure of the inclination angle of the system, confirming a low inclination of ${\rm 31^{+9}_{-5} \, degrees}$. We report a subsolar iron abundance both in the accretion disk and in the ISM, which is likely related to the source location in the metal-poor globular cluster NGC 6624. Moreover, the presence of the helium white dwarf companion could further impact the metallicity of the accretion disk. 
 
The polarimetric analysis confirms a clear trend already detected by \cite{DiMarco.4U1820}, with the PD increasing from an upper limit of $\sim 1.2\%$ at lower energies (2--4 keV) to $8.0\% \pm 3.6\%$ in the 7--8 keV band. This behavior is extremely peculiar and, despite reflected photons result being highly polarized (as expected), their contribution is not sufficient to explain the high PD observed at high energies, especially considering the relatively low inclination of the system. Therefore, to explain the high PD at
high energies, an additional contribution to the polarization would be required or a specific shape of the Comptonizing region that would differ from the classical boundary+spreading layer.  

\begin{acknowledgements}  AM is supported by the European Research Council (ERC) under the European Union's Horizon 2020 research and innovation programme (ERC Consolidator Grant ``MAGNESIA'' No. 817661, PI: Rea) and by the grant SGR2021-01269 from the Catalan Government (PI: Graber/Rea).  
The authors acknowledge financial support from PRIN-INAF 2019 with the project ``Probing the geometry of accretion: from theory to observations'' (PI: Belloni). WL conducted this research during, and with the support of, the Italian national inter-university PhD program in Space Science and Technology. 
AG, SB, FC, SF, GM, AT, and FU acknowledge financial support by the Italian Space Agency (Agenzia Spaziale Italiana, ASI) through the contract ASI-INAF-2022-19-HH.0. This research was also supported by the Istituto Nazionale di Astrofisica (INAF) grant 1.05.23.05.06: ``Spin and Geometry in accreting X-ray binaries: The first multi frequency spectro-polarimetric campaign''. SF has been supported by the project PRIN 2022 - 2022LWPEXW - ``An X-ray view of compact objects in polarized light'', CUP C53D23001180006.
ADM contribution is supported by the Istituto Nazionale di Astrofisica (INAF) and the Italian Space Agency (Agenzia Spaziale Italiana, ASI) through contract ASI-INAF-2022-19-HH.0.

\end{acknowledgements}
\bibliographystyle{aa}
\bibliography{biblio}
\end{document}